\documentclass[twocolumn]{aastex6}

\usepackage{wrapfig,lipsum,booktabs}
\usepackage{natbib}
\usepackage{silence}

\shorttitle{X-Ray Variability in RZ2109}
\shortauthors{Dage et al.}

\begin{document}

\title{X-ray variability from the ultraluminous black hole candidate X-ray binary in the globular cluster RZ 2109}
\author{Kristen C. Dage, \altaffilmark{1}  Stephen E. Zepf, \altaffilmark{1} \\
Arash Bahramian, \altaffilmark{1} Arunav Kundu, \altaffilmark{2}  Thomas J. Maccarone,  \altaffilmark{3} Mark B. Peacock \altaffilmark{1}}

\altaffiltext{1}{Department  of  Physics  and  Astronomy,  Michigan  State  University,  East
Lansing, MI 48824}
\altaffiltext{2}{Eureka Scientific, Inc., 2452 Delmer Street, Suite 100 Oakland, CA 94602, USA}
\altaffiltext{3}{Department of Physics, Box 41051, Science Building, Texas Tech University, Lubbock, TX 79409-1051, USA}
\email{kcdage@msu.edu}

\begin{abstract}
We present the results of long-term monitoring of the X-ray emission from the
ultraluminous X-ray source XMMUJ122939.9+075333 in the extragalactic globular 
cluster RZ2109. The combination of the high X-ray luminosity, short 
term X-ray variability, X-ray spectrum, and optical emission suggest that this system is likely an accreting black hole in a globular cluster. To study the long-term behavior of the  X-ray emission from this source, we analyze both new and archival \textit{Chandra} and \textit{XMM-Newton} observations, covering 16  years from 2000 to 2016. For all of these observations, we fit extracted spectra of RZ2109 with \textsc{xspec} models. The spectra are all dominated by a soft component, which is very soft with typical fit temperatures of T $\simeq$ 0.15 keV. The resulting X-ray fluxes show strong variability on short and long timescales. We also find that the X-ray spectrum often shows no significant change even with luminosity changes as large as a factor of five.

\end{abstract}

% http://journals.aas.org/authors/keywords2013.html

\section{INTRODUCTION}
\label{introduction}

Determining whether or not globular clusters host black holes, and if so 
how many and what masses has been the subject of both longstanding interest 
% decided it need a ``the subject''
and current theoretical work (e.g. \citealt{spitz1969} to \citealt{chatterjee17}).
This interest has grown dramatically with the detection of merging 
black holes by LIGO \citep{abbottprl}, as black hole mergers in globular 
clusters are one of the leading possibilities for the origin of the LIGO 
sources  \cite[e.g,][]{abbottapj, 2016PhRvD..93h4029R}.
There is little question black holes are present early in the life of 
a globular clusters - hundreds to thousands of stellar mass black holes 
are expected to be produced in typical globular clusters as the result 
of standard stellar
evolution \cite[e.g.,][]{ivanova2010}. Early work suggested that dynamical 
interactions among the black holes may
eject many or almost all of them (e.g. \citealt{kulkarni1993}, \citealt{1993Natur.364..423S}). However,
recent simulations have found that this process is not as efficient as earlier expectations and current models generally
predict the retention of a significant number of black holes in globular clusters (e.g. \citealt{2015ApJ...800....9M}, \citealt{heggie14}, \citealt{2013MNRAS.430L..30S}).

From an observational perspective, some of the first and strongest
evidence for the presence of black holes in globular clusters has come from
X-ray and optical studies of extragalactic globular clusters.
One way to find black hole candidates in globular clusters
is to identify globular cluster X-ray sources with luminosities in excess 
of the Eddington limit for an accreting neutron star, which may be 
indicative of a more massive black hole primary.
%change here - was ``suggested.''
No such source is found in Galactic globular clusters.
However, over the very large
sample of extragalactic globular clusters a number of such sources
have now been identified in Chandra observations of early-type galaxies
in the local universe. Because extragalactic globular clusters are 
unresolved in the X-rays, short term variability in these ultraluminous 
X-ray sources (ULXs) is 
critical for eliminating the possibility that the high X-ray luminosity 
arises from the superposition of multiple accreting neutron stars 
\cite[e.g.,][]{kalogera2004}. 
A handful of such high L$_X$, variable
sources are now known, starting with RZ2109 \citep{2007Natur.445..183M}, 
making these among the best candidates for accreting black holes 
in globular clusters (e.g. \citet{2012ApJ...760..135R}, 
\citet{2011MNRAS.410.1655M}, \citet{irwin2010} and references therein). 

Additional information about these ultraluminous sources can help 
constrain the nature of the accreting system. For example the
source in RZ2109 has strong [OIII]4959,5007 emission with a velocity
width of several thousand km/s and absolutely no hydrogen observed in 
emission \citep{2008ApJ...683L.139Z}. The strong presence of 
[OIII]5007 and absence of H$\beta$ is indicative of accretion from 
a very O-rich and H-poor donor, for which a CO white dwarf seems 
most likely \citep{2014ApJ...785..147S}.
The broad velocity and high luminosity of the outflow is also consistent 
with outflows from sources accreting around their Eddington limit, 
based on both empirical and theoretical 
work \cite[and references therein]{2008ApJ...683L.139Z}
Moreover, the [OIII]5007 can not be strongly beamed. Therefore, the lack of 
similar [OIII] emission in objects without high L$_X$ places valuable 
limits on any beaming of the X-rays in the RZ2109 source.

It is also interesting to compare the ULXs discovered in globular clusters
with the more widely known ULXs in star forming galaxies (see \citet{2017ARA&A..55..303K} for a review of the latter). While sharing
the property of high X-ray luminosity, there are many differences between
these populations. The donor stars in the star forming ULX systems are
typically high mass stars, while such stars are long dead in globular 
clusters, and the donor stars in globular cluster systems are either 
white dwarfs \cite[as in RZ2109,][]{2014ApJ...785..147S}
or other lower mass stars. Moreover, for a ULX in a star forming galaxy the 
accreting compact object will have been recently formed, and in the 
case of a neutron star may have an extremely high magnetic field. 
In contrast, globular clusters have populations of compact objects that
formed long ago which can make close binary systems through dynamical
interactions within the globular cluster
\citep{ivanova2010}. 

These underlying physical differences may be matched to observational
differences between the star forming and globular cluster ULXs. One
of the most striking results in the study of ULXs in star forming
galaxies is that at least some of the ULXs have accretors that are
neutron stars rather than black holes and are thus accreting at many
times their formal Eddington limit (e.g. \citealt{2014Natur.514..202B}, \citealt{2016ApJ...831L..14F}, \citealt{2017MNRAS.466L..48I}, \citealt{2017Sci...355..817I}).
Models to account for these generally involve some combination of 
extremely large magnetic fields and beaming (review by \citet{2017ARA&A..55..303K}
and references therein). Because most of these models
predict that pulses will not be observed at all times and depend on various
geometries, it is possible that many ULXs in star forming galaxies
are neutron star accretors (\citealt{2017MNRAS.468L..59K}, 
\citealt{2017MNRAS.470L..69M}).

The likely absence of both extreme magnetic fields and substantial 
beaming differentiates globular cluster ULX sources like RZ2109 from some
star forming ULXs and supports identifying the RZ2109 
accretor as a black hole \citep{2012ApJ...759..126P}. 
However, studies of ULXs in star forming galaxies provide an extensive 
set of phenomenological and theoretical work on super-Eddington accretion 
onto compact objects and its observational manifestations to which 
observations of RZ2109 can be compared 
(e.g. \citealt{2007MNRAS.377.1187P}, \citealt{2009MNRAS.397.1836G},
\citealt{2013MNRAS.435.1758S}, \citealt{2015MNRAS.454.3134M}). Broadly speaking 
these papers relate accretion rate relative to Eddington and viewing angle
to observed properties such as X-ray spectrum, luminosity, and possible
variability based on various assumptions about the underlying astrophysics
\cite[see review by][]{2017ARA&A..55..303K}. 

The goal of this paper is to analyze the now large number of X-ray observations
of RZ2109 over 16 years, with the aim of constraining the nature
of this likely accreting black hole system. RZ2109 is a very well-studied
system, with extensive optical spectroscopy 
(\citet{2014ApJ...785..147S}, \citet{2011ApJ...739...95S},
\citet{2008ApJ...683L.139Z} and references therein), and several extant X-ray
studies utilizing Chandra and XMM-Newton. Here we report multiple new Chandra 
and XMM-Newton observations, and combine these with archival data to study 
the variability 
of the X-ray emission from RZ2109 over a broad range of time scales. The paper
is arranged so that the observations are presented in Section \ref{sec:observations}, the results
from the analysis of these observation in Section \ref{sec:results}, and the conclusions in Section \ref{sec:conclusions}.

\section{DATA AND ANALYSIS}
\label{sec:observations}

RZ2109 has been observed numerous times by \textit{XMM-Newton} and \textit{Chandra} observatories over the last 16 years. We reduced and analyzed all of these observations as tabulated in Table \ref{table:chandraobsinfo}.
The background flare filtered (see further in text for details) \textit{XMM-Newton} net count rates were obtained by filtering the energy in the range 0.3-10 keV in the spectral extraction.  We then loaded the spectra into \textsc{XSPEC}, and obtained the net count rates (with no model) by using the show rate command.
 The \textit{Chandra} source count rates were calculated using funtools \footnote{https://github.com/ericmandel/funtools}  to list the counts in the source and background areas based on an image filtered in the 0.3-10 keV range. We then background subtracted the average source counts and divided by the observation length. 

\begin{table}[h]
\centering
\caption{\label{table:chandraobsinfo}  \textit{Chandra} and background flare filtered \textit{XMM-Newton} observations, background subtracted average source count rates and raw  source counts in 0.3-10 keV. } 
\begin{tabular}{lccccc}
\hline
\hline
ObsID     &Date  & Exposure & Avg. Rate & Src. Counts\\ 
&&(ks)& count/s \\
\hline
322  \footnote{This observation and below: \textit{Chandra}}  & 2000-03-19 & 10            & 4.3 $\times 10^{-3}$    &48 \\ 
321       & 2000-06-12 & 40            & 8.7$\times 10^{-3}$  &398   \\ 
8095 & 2008-02-23 & 5 & 1.1 $\times 10^{-2}$ &60  \\
11274     & 2010-02-27 & 40            & 2.1$\times 10^{-4}$ &19  \\ 
12978    & 2010-11-20 & 20            & 9.6$\times 10^{-4}$  &19   \\ 
12889  & 2011-02-14 & 140           &    2.4$\times 10^{-3}$  &425 \\ 
12888   & 2011-02-21 & 160           &     1.0$\times 10^{-3}$ &230   \\ 
16260 & 2014-08-04 & 25            & 3.1$\times 10^{-3}$  &79 \\ 
16261     & 2015-02-24 & 25            & 9.0$\times 10^{-5}$ &3   \\ 
16262  & 2016-04-30 & 25            & 5.6$\times 10^{-4}$  &16  \\ \hline
0112550601\footnote{ This observation and below: \textit{XMM-Newton} }	    & 2002-06-05 & 11            & 2.3 $\times 10^{-2}$ &282  \\ 
0200130101       & 2004-01-01 & 72      & 4.5 $\times 10^{-3}$  &465\\ 
0761630101 & 2016-01-05 & 44& 2.1$\times 10^{-2}$ &1147 \\
0761630201 &2016-01-07& 35& 4.3 $\times 10^{-4}$  &29\\
0761630301 &2016-01-09&65& 3.9$\times 10^{-4}$& 217 \\

\hline
% SOURCE: /home/patricio/ast/esp01/analyses/ligtcurve/attempt_547/results.txt
\end{tabular}
\end{table}

\subsection{Observations}

	The \textit{Chandra}  observations include both three new datasets we obtained in 2014, 2015, and 2016, and archival data going back to 2000. We used \textsc{ciao} version 4.9 \footnote{http://cxc.harvard.edu/ciao/}  \citep{2006SPIE.6270E..1VF} for analysis of all \textit{Chandra} data. For most on-axis observations, we manually extracted the spectrum from the source and background regions using \texttt{specextract}. For off-axis observations we used \textsc{acis-extract} \footnote{http://www2.astro.psu.edu/xray/docs/TARA/ae$\_$users$\_$guide.html} \citep{2012ascl.soft03001B} to extract the source regions \citep{2012ascl.soft03001B}. Specifically, in observations 12888 and 12889, the source is located on ACIS chip 8 and thus far off-axis. Additionally in observation 12888, it is located on the edge of this chip and is affected by dithering and edge effects. Given these issues and low signal-to-noise ratio of the detection in these observations, we used \textsc{acis-extract}  to extract the spectra. Source extraction regions constructed by \textsc{acis-extract} are polygons approximating \textit{Chandra}-ACIS point spread function based on MARX \citep{2012SPIE.8443E..1AD} simulations\footnote{We note that MARX 5.0, 5.1, and 5.2 simulate PSF of off-axis sources inaccurately (see https://github.com/Chandra-MARX/marx/pull/21). We have used MARX 5.3.2 for this work, which has addressed this issue.}. \textsc{acis-extract} also applies PSF corrections to ancillary response files (ARF) and exposure and background scaling corrections to the spectrum to take into account edge effects. 

RZ2109 was also observed with \textit{XMM-Newton} in 2002, 2004, 2008\footnote{Observation 0510011501 did not have enough information left post-background flare filtering and thus is not used in this analysis.}, and three times in 2016 (see Table \ref{table:chandraobsinfo}). We used  \textsc{sas} 16.1.0\footnote{https://www.cosmos.esa.int/web/xmm-newton/sas} to extract  the spectra from the MOS1, MOS2 and pn detections. 

We set FLAG$==0$ to screen conservatively \footnote{https://heasarc.gsfc.nasa.gov/docs/xmm/hera\_guide/node33.html}, and  originally extracted single and double events (pattern $<= 4$) as recommended for \textit{XMM} pn, however, in observations heavily impacted with flares (0200130101, 0761630201, and 0761630301), we only extracted single events (pattern $==$ 0). For MOS1 and MOS2, we select (pattern$<=$12). The data was filtered for high background flares by only selecting times at which the background was constant. Those times were determined by examining the background light curve from the PPS. We ignored any counts below 0.2 keV.  To account for differences among the three detectors when fitting, a constant factor was added to the best fit models; the value for pn was frozen at 1.0, while the values for MOS1 and MOS2 were free.

 We used  \textsc{xspec} version 12.9.1\footnote{https://heasarc.gsfc.nasa.gov/xanadu/xspec/} \citep{1996ASPC..101...17A} to analyze the X-ray spectra of both new and archival  \textit{Chandra} and \textit{XMM-Newton} observations. All \textsc{xspec} analysis used the abundance of elements from \citet{2000ApJ...542..914W}. 

We used the \textit{F-test} function to compare the $\chi^2$ statistics of a single  disk component model with an absorption term to a two component model (disk component added to a powerlaw model pegged from 0.5-8 keV, also with absorption) for the three \textit{Chandra} observations with the highest counts (321, 12888, and 12889). The probability that the improved fit statistics of the two component model is due to chance is respectively: 0.002, 0.006, 1.4e-05. Similarly, the \textit{F-test} probabilites of a single powerlaw model with absorption compared  to the absorbed two component model are: 0.047, 0.047 and 0.001. Therefore, we fit a multicolor disk (MCD) model (\texttt{diskbb}) added to  a power law \texttt{pegpwrlw}, and multiplied by the absorption component to all observations. We note for completeness that if a power law is fit as a single-component model to the \textit{Chandra} data, its index is in the range from 3.2 to 4.3, while the \textit{XMM-Newton} data typically have a single-component powerlaw index around 3.0.  
 In all our fits, we include an absorption term, \texttt{tbabs}, fixed to a foreground hydrogen column density of N$_H = 1.6 \times 10^{20}$  cm$^{-2}$ \footnote{http://cxc.harvard.edu/toolkit/colden.jsp}.  We found no evidence for a second absorption column; we fit the highest count \textit{Chandra} data with a second absorption parameter and found that in each case, the best fit value was consistent with zero. 

For the bulk of the observations $\chi^2$ was used as the fitting statistic. Spectra with more than 100 source counts were binned in groups of 20;  spectra with fewer counts than that were binned with 1 count per bin  and fit with  c-stats \citep{1979ApJ...228..939C}. 
% !!! Could put the chi2 stat stuff here. !!!

To estimate the unabsorbed fluxes in the 0.5-8keV range we used \textsc{xspec}'s multiplicative model \texttt{cflux} \footnote{https://heasarc.gsfc.nasa.gov/xanadu/xspec/manual/XSmodelCflux.html}  with the best fit spectral model. 
After adding in the cflux component and refitting, we then used  the  error command in  \textsc{xspec} on the flux parameter and obtained upper and lower bounds on the fluxes of each observation to the 90\% confidence interval. All parameter errors were also obtained in this manner.

While all of our fitting is carried out in the 0.5-8keV range appropriate for \textit{Chandra} data, many X-ray results are given in the 0.2-10 keV range. Therefore to compare to other work, we calculate the 0.2-10 keV fluxes and luminosities based on our spectral fits to the  data from 0.5-8 keV, also using \texttt{cflux}.

Tables  \ref{chpar}  and \ref{xmmpar} show best fit parameters, fit statistics and the fitted flux for \textit{Chandra} and \textit{XMM-Newton} observations respectively. The fluxes and fit parameters are also plotted in Figure \ref{fig:params}.

\begin{table*}[h]
\caption{\textit{Chandra} Fit Parameters and Fluxes (0.5-8 keV) for \textsc{xspec} best fit model \texttt{tbabs*(diskbb+pegpwrlw)}. Hydrogen column density ($N_H$) frozen to 1.6 $\times 10^{20}$  cm$^{-2}$. }
\label{chpar}
\begin{tabular}{lclclclclclclclclc}

\hline
\hline
 Date &$T_{in}$  & Disk Norm\footnote{$(R_{in}/D_{10})^2 \textrm{cos}\theta$ }&  $\Gamma$&  Powerlaw Flux&  $\chi^2_{\nu}$/d.o.f. \footnote{Reduced $\chi^2$ per degree of freedom} & Unabsorbed  Flux\\
&(keV)&&&(erg cm$^2$ s $^{-1})$&&(erg cm$^2$ s$^{-1}$) \\ \hline
2000-03-19& 0.11 $ _{-0.03} ^{+0.04}$&107 $^{+670}_{-90}$& 1.3 $ _{-1.1}  ^{+1.2}$&3.0($_{-1.4}^{+2.5}$)$\times$$10^{-14}$& N/A  \footnote{Too few data points for $\chi^2$ statistics. Pearson $\chi^2$:   80.63 using 41 PHA bins.}   & 5.7($_{-1.6}^{+2.4}$) $\times$$10^{-14}$ \\ 
2000-06-12 &  0.12 $\pm 0.02$ &68$^{+85}_{-34}$&1.7 $\pm 0.5 $ &3.8($\pm 0.8)$$\times$$10^{-14}$& 3.14/16 & 7.0 ($\pm 0.9)$ $\times$$10^{-14}$\\
2008-02-23 & 0.13 $ _{-0.02} ^{+0.03} $ & 60$ _{-42} ^{+164} $ & 0.5  $_{-2.5}  ^{+1.4}  $& 2.2($_{-1.8}^{+2.7}$)$\times$$10^{-14} $& N/A \footnote{Too few data points for $\chi^2$ statistics. Pearson $\chi^2$:  570.25 using 44 PHA bins.} & 4.5($_{-1.0}^{+1.2}$)$\times 10^{-14}$ \\
 2010-02-27 \footnote{Count rate from \citet{2010MNRAS.409L..84M}, fit with \textsc{pimms}. } &---   & --- &(3.5) \footnote{Best fit powerlaw index from single component powerlaw model.}    & 9.2 ($^{+10.0}_{-6.2})$ $\times$$10^{-16}$ &---  & 9.2($^{+10.0}_{-6.2})$ $\times$$10^{-16}$ \\ 
 2010-11-20 & 0.13 $_{-0.04} ^{+0.08}$ & 3.1$ _{-2.9} ^{+38.3} $&0.2$^{+1.6}_{-2.1}$&6.7 ($^{+10.0}_{-4.5})$ $\times$$10^{-15} $ & N/A \footnote{Too few data points for $\chi^2$ statistics. Pearson $\chi^2$:  20.99 using 22 PHA bins.} &  9.6($_{-4.6}^{+8.8}$)$\times$ $10^{-15}$\\ 
2011-02-14& 0.13 $ \pm 0.02 $ & 21 $_{-14}^{+36}$ &1.3 $\pm 0.6$ &1.9 ($\pm 0.4$ )$\times$$10^{-14}$& 1.19/16&
3.6($\pm 0.4$)$\times$$10^{-14}$  \\ 
2011-02-21 & 0.11 $\pm 0.03 $ &30 $_{-23}^{+135}$& 1.8 $ \pm 0.6$ &1.0 ($\pm 0.2$ )$\times$$10^{-14}$& 1.69/7  &1.6($\pm 0.3)$$\times$$10^{-14}$  \\ 
 2014-08-04 & 0.12 $ _{-0.03} ^{+0.04} $&39 $^{+233}_{-34}$  & 2.0 $ \pm 0.7$ & 2.6($\pm 0.8$)$\times$$10^{-14}$& N/A \footnote{Too few data points for $\chi^2$ statistics. Pearson $\chi^2$: 86.80 using 64 PHA bins.}&2.8($_{-0.5}^{+0.6}$)$\times$$10^{-14}$\\ 
 2015-02-24 \footnote{Marginal detection--fit with \textsc{pimms}.} & --- &---  & (3.5)  & 9.3($^{+27.7}_{-7.9})$$\times$$10^{-16}$ & --- &    9.3($^{+27.7}_{-7.9})$$\times$$10^{-16}$\\ 
 2016-04-30 & 0.09 $\pm 0.05$ &242 $_{-239}^{2169}$& 1.8$_{-1.6} ^{ +1.5}$  &6.9($_{-3.5}^{+5.0}$)$\times$$10^{-15}$& N/A \footnote{Too few data points for $\chi^2$ statistics. Pearson $\chi^2$: 12.07 using 16 PHA bins.} &2.0($^{+1.1}_{-0.8})$$\times$$10^{-14}$\\ 
\hline
\end{tabular}
\end{table*}

Two \textit{Chandra} observations have extremely low average source counts. \citet{2010MNRAS.409L..84M}  have previously found 19 source count rates with a background of 10.8 counts for obsID 11274, which, despite being highly off-axis,  is significant at the 95\% confidence level  \citep{1986ApJ...303..336G}.  In obsID 16261, we detect three counts in the source region, which - considering the expected scaled background of 1 count - is also significant at the 95\% confidence level. To estimate a flux for both of these observations, we took the background subtracted count rates and used \textsc{pimms} \footnote{http://cxc.harvard.edu/toolkit/pimms.jsp} to fit with a powerlaw index of 3.5, which was the common best fit to the single powerlaw model of the other \textit{Chandra} data. \textit{XMM-Newton} observation 0761630201 had very few counts left post background flare filtering. This, in conjunction with a relatively high background ($\simeq$ 50\%) meant that detailed spectral analysis was not possible. However, we were able to fit a single component disk model to the data and obtain a flux using \texttt{cflux}. This lends significant uncertainty to this flux estimate. 

% !!! not sure this paragraph belongs here at the bottom, but oh well. !!!

%\begin{table}[h]
%\centering
%\caption{\label{table:xmmobsinfo} \textit{XMM-Newton} Observations and average source count rates drawn from \textit{XMM-Newton} Pipeline Processing System (PPS).While  RZ2109 was strongly detected in observation 0761630301, no count rate was given for it. }
%\begin{tabular}{lclclc}
%\hline
%\hline
%DetID     &  Date  & Exposure  & Catalog Avg. Rate \\ 
%& & (ks) & count/s \\
%\hline
%0112550601	     & 2002-06-05 & 15            & 1.2 $\times 10^{-1}$ & XMM  \\ 
%0200130101       & 2004-01-01 & 83       & 1.5$\times 10^{-2}$  & XMM\\ 
%0510011501	 & 2008-01-07 & 1.3         & 2.0 $\times 10^{-2}$	 & XMM     \\ 
%0761630101 & 2016-01-05 & 116& 3.7$\times 10^{-3}$ & XMM \\
%0761630201 &2016-01-07& 116.5& 1.9 $\times 10^{-3}$  & XMM \\
%0761630301 &2016-01-19&117& 4.2 $\times 10^{-3}$  & XMM \\

%\hline
% SOURCE: /home/patricio/ast/esp01/analyses/ligtcurve/attempt_547/results.txt
%\end{tabular}
%\end{table}

\begin{table*}[h]

\caption{\textit{XMM-Newton} Fit Parameters and Fluxes (0.5-8 keV) for \textsc{xspec} model \texttt{tbabs*(diskbb+pegpwrlw)}. Hydrogen column density ($N_H$) frozen to 1.6 $\times 10^{20}$  cm$^{-2}$.  }
\label{xmmpar}

\begin{tabular}{lclclclclclclclclc}
\hline
\hline
 Date & CCF Constant &$T_{in}$  & Disk Norm\footnote{$(R_{in}/D_{10})^2 \textrm{cos}\theta$} &  $\Gamma$ &  Powerlaw Flux & $\chi^2_{\nu}$/d.o.f. \footnote{Reduced $\chi^2$ per degree of freedom}  & Unabsorbed Flux \\
&pn/MOS1/MOS2&(keV)&&&(erg cm$^2$ s $^{-1}$)&&(erg cm$^2$ s$^{-1}$) \\ \hline
 2002-06-05 &(1.0)/1.20/0.83& 0.14  $_{-0.02}^{+0.06}$&18$_{-16}^{+25}$& 2.6 $ _{-3.9} ^{+0.7} $&  3.9 ($\pm$ 1.5)$\times$$10^{-14}$&1.73/16 & 6.3 ($^{+3.8}_{-1.0}$)$\times$$10^{-14}$\\ 
 2004-01-01 \footnote{PN detection started ~3ks after MOS detectors}&(1.0)/1.57/1.59& 0.17  $\pm 0.02$& 1.8 $_{-0.5}^{+0.8}$& 0.6  $_{ -0.9} ^{+0.7}$ &9.2($_{-3.4}^{+3.7}$)$\times$$10^{-15}$&2.31/45&1.7 ($\pm 0.4$)$\times$$10^{-14}$ \\ 
 2016-01-05 &(1.0)/0.85/1.08 &0.16$ \pm 0.01$ &7.2 $_{-1.8}^{+2.6}$& 1.0  $\pm 0.4$ &2.7 ($\pm$ 0.5)$\times$$10^{-14}$& 1.38/56& 5.0 ($\pm 0.6$) $\times$$10^{-14}$\\ 
 2016-01-07 &---&---- &--- &  ---&--- & --- &  8.9($^{+13.0}_{-8.2}$)$\times$$10^{-16}$\\ 
 2016-01-09 & N/A \footnote{Too few counts in MOS1 \& MOS2, used pn only}&0.15$ _{-0.02} ^{+0.15} $&0.2 $_{-0.1}^{+1.4}$& 1.4 $_{-0.6}^{+1.4}$&$\leq$7.0$\times$$10^{-15}$& 0.59/6& 2.3($^{+5.2}_{-2.2}$)$\times$$10^{-15}$\\ 
\hline
\end{tabular}
\end{table*}

\section{RESULTS}
\label{sec:results}

\begin{figure}[h]
\includegraphics[width=9cm]{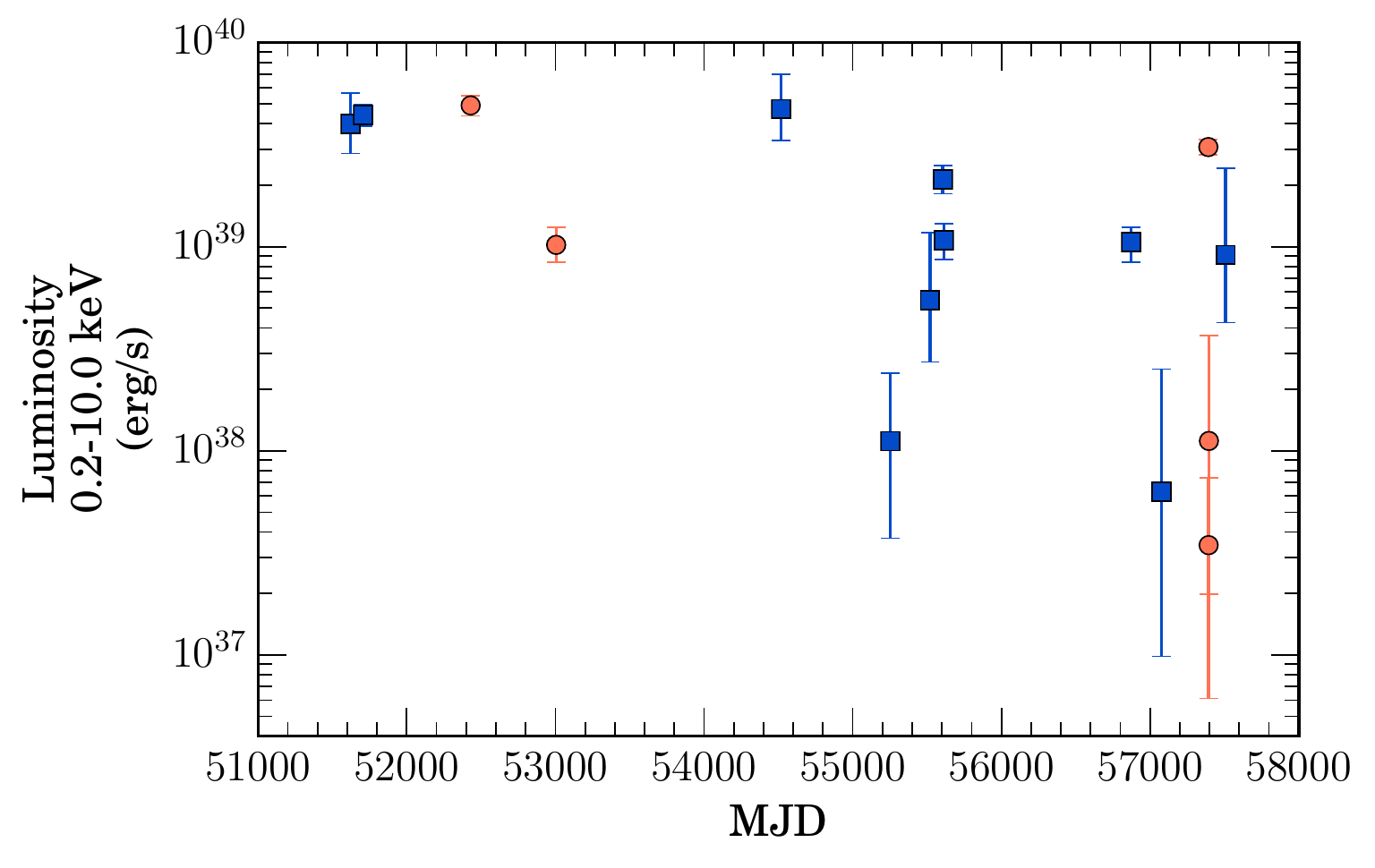}
  \caption{Luminosity variability of RZ2109 in energy band 0.2-10.0 keV. \textit{Chandra} data is represented by blue squares, and the \textit{XMM-Newton} data by orange circles.  }

  \label{fig:fluxvar}
\end{figure}

The overall goal of this paper is to monitor variations in the X-ray luminosity of RZ2109 over the time covered by all of the available data, ranging from 2000 to 2016. Figure \ref{fig:fluxvar} shows the luminosities in the 0.2-10 keV range, which were calculated using the fluxes from  Section \ref{sec:observations} and a distance of 16.1 Mpc \citep{macri}. The luminosities are also listed in Table \ref{table:all-lum}.  
 
One of the main results apparent from Figure \ref{fig:fluxvar} and Table \ref{table:all-lum} is that RZ2109 varies significantly over all of the time scales observed, from days to years. During some observations RZ2109 is observed to have L$_X \sim 4 \times 10^{39}$ erg s$^{-1}$, while at other times it is observed to have L$_X \sim 2-3 \times 10^{38}$ erg s$^{-1}$ or even fainter, along with various times at which RZ2109 is found to be between these luminosities. This variability is surprising because there is strong evidence in other ways that the source is a stellar mass black hole accreting material from a carbon-oxygen white dwarf at a very high rate, at or somewhat above its formal Eddington limit \cite[e.g.,][]{peacock2012a, 2012ApJ...760..135R}. In such a case the source is expected to be persistent because accretion disks in high luminosity, short period
ultracompacts like this are not expected to have ionization instabilities
\cite[e.g.,][]{2010MNRAS.406.2087M}. While there are beginning to be counter-examples to this argument \cite[see][]{2010MNRAS.406.2087M}, understanding such systems may give important clues to the formation and accretion processes
in these globular cluster black hole sources. 
It is interesting to compare the variability of RZ2109 to that of other ultraluminous X-ray sources in extragalactic globular clusters. Of the six such sources published -  \citet{2007Natur.445..183M}, \citet{2010ApJ...725.1805B}; \citet{2010ApJ...721..323S}; \citet{irwin2010}; \citet{2011MNRAS.410.1655M}; \citet{2012ApJ...760..135R}, all vary, with at least three of them varying by more than an order of magnitude. There is a need to be careful about variability in this list, because variability is also one of the criteria to ensure that most of the X-ray flux comes from a single source and not multiple sources in the globular cluster, and variability is one of the criteria used in these papers. However, in these cases, the ultraluminous sources are typically the brightest X-ray sources among the globular cluster sources in each galaxy. So substantial and large variability appears to be the norm for ultraluminous X-ray sources in extragalactic globular clusters.

\begin{figure}[h]
\includegraphics[width=9cm]{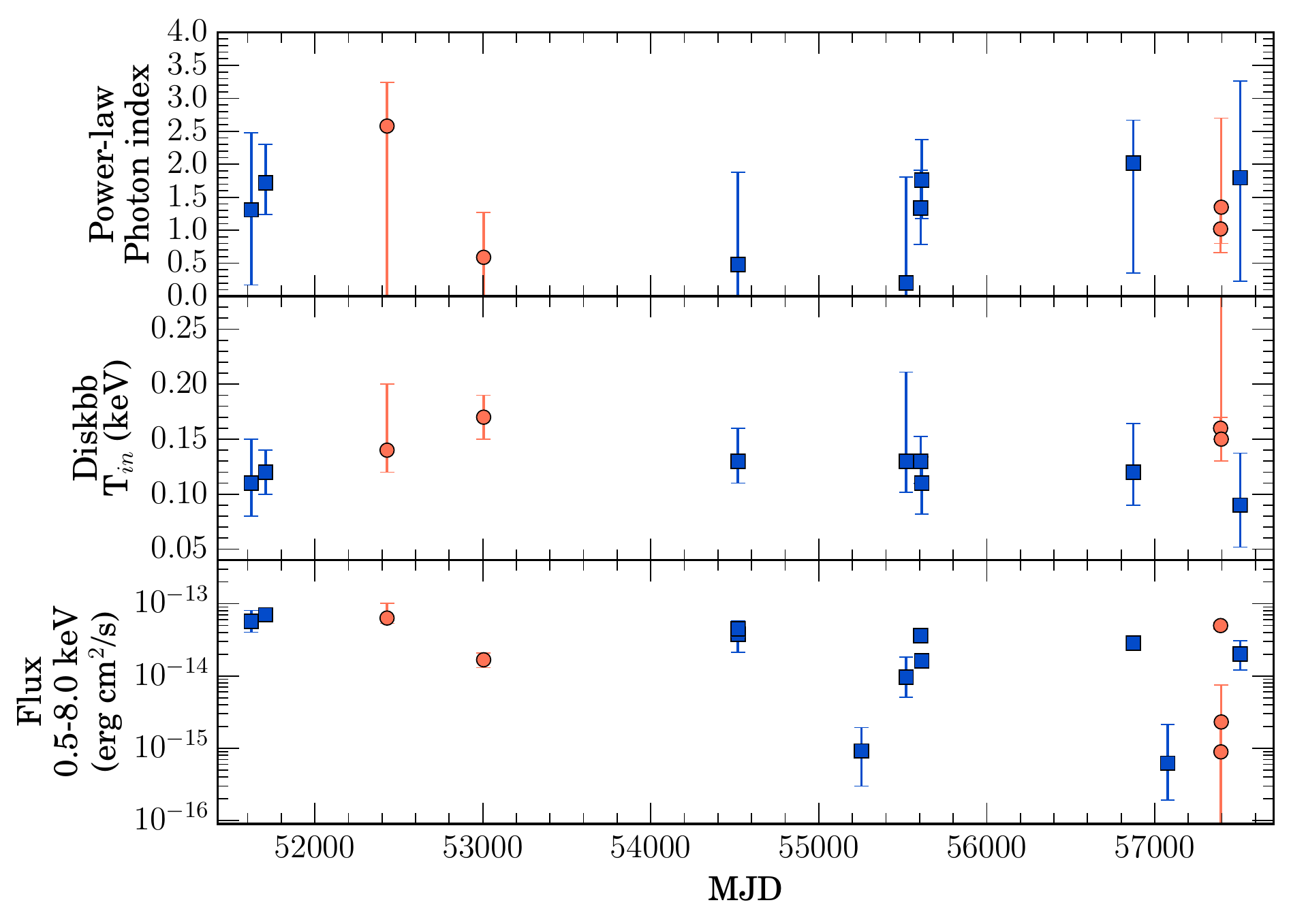}
  \caption{Top panel shows best fit powerlaw index for \textit{Chandra} and \textit{XMM} observations, middle panel shows disk temperature,  lowest panel shows calculated unabsorbed model flux.  \textit{Chandra} data is represented by blue squares, and the \textit{XMM-Newton} data by orange circles.  All parameters and fluxes are in the 0.5-8 keV band. }

  \label{fig:params}
\end{figure}

Given the  luminosity variability observed in these sources, it is natural to test whether there is any corresponding spectral variability. A key feature of the variability RZ2109 found here is that there is often {\it no} evidence for corresponding changes in the X-ray spectra of RZ2109. The similarity of the X-ray spectra at different observed fluxes can  be seen  in Tables \ref{chpar} and \ref{xmmpar} and Figure \ref{fig:params}. This result is different than that found in the original variability discovery for RZ2109 which showed that the decrease in flux seen within the 2002 \textit{XMM-Newton} observation was driven by a decrease in the soft component that can be interpreted as a change in the column density of absorbing material along the line of sight (\citealt{2007Natur.445..183M}, \citealt{2008MNRAS.386.2075S}). While this is still true for the flux change within the 2002 \textit{XMM-Newton} observation, such a model can not explain most of the variability among the many observations shown here. Other ULX sources in extragalactic globular clusters show a range of behavior in the relationship between luminosity and spectral variability. \citet{2010ApJ...721..323S} find a ULX in an extragalactic globular cluster in NGC 1399 with more than a factor of ten decrease in luminosity and no evident change in spectral shape. 
On the other hand, a different extragalactic globular cluster ULX in NGC 4649 studied by \citet{2012ApJ...760..135R} does exhibit spectral changes in some observations. The overall picture is that some spectral variability happens, but there are clear observations in multiple sources of little or no spectral variability even with order of magnitude luminosity changes.

It is also natural to ask whether there is any overall longterm trend of L$_X$ with time for RZ2109. Unfortunately the data are not quite adequate for addressing this question specifically. It is intriguing that many of the fainter fluxes appear to be found at more recent times. However, there are observations in 2014 and 2016 in which L$_X \simeq 3 \times 10^{39}$ erg s$^{-1}$, and thus well within the ULX regime and not much different than observations in 2000 and 2002. Given the short term variability clearly evident in RZ2109, one way to get at the long-term changes in RZ2109 may be to study its [OIII]5007 emission which appears to be emitted over a region of light months to light years, and therefore samples the overall photoionizing flux averaged over those timescales \citep{2012ApJ...759..126P}. The physical origin of the variability RZ2109 has not yet been established. As noted above, the accretion disk is not expected to be subject to ionization instabilities, so a different mechanism must be operating. 

One natural mechanism to produce changes in the accretion rate over time is to have small changes in the eccentricity of the orbit \cite[e.g.,][]{1984ApJ...284..675H}, perhaps due to the Kozai mechanism \citep{1962AJ.....67..591K}, in which a lower mass third star in an outer orbit introduces eccentricity into the main two body system. For typical parameters for an RZ2109-like system and a mass of the third star of 0.5$M_\odot$, the Kozai timescale is roughly a decade. This therefore may account for any long term trend we may see in RZ2109, but not very short term variability. This is why it was attractive originally to try to explain RZ2109's variability with varying absorption, but such an explanation fails to account for the absence of spectral changes seen in many datasets since. Other possibilities for short-term variability are listed in the \citet{2010MNRAS.406.2087M}  study of similar variability of the Galactic ultracompact binary 4U 0513$-$40. These  possibilities include tidal disc instabilities \cite[e.g.,][]{1988MNRAS.232...35W, 1995PASJ...47L..11O} and irradiation of the donor star leading to modulations of the accretion rate \cite[e.g.,][]{1986AA...162...71H}. It is not yet clear whether these can account for variability observed in RZ2109.

The observed X-ray spectrum and variability of RZ2109 can also be 
compared to the well-known ULXs observed in star forming galaxies 
\citep{2017ARA&A..55..303K}.
Compared to most of these ULXs not in globular clusters, RZ2109 is
significantly softer and much more variable. Within the field ULX
population, there is a class of objects known as ultrasoft ULXs (ULSs)
that are both softer and more variable than most ULXs
\cite[e.g.,][]{2017MNRAS.467.2690E, 2016MNRAS.456.1859U}.
These papers also find that the variability in ULSs is primarily
at higher X-ray energies, although there is an exception to this
general characteristic \citep{2016ApJ...831..117F}.
Thus RZ2109 differs from most field ULSs in that its variability 
is either mostly at low energies \cite[e.g.,][]{2008MNRAS.386.2075S},
or the luminosity varies equally at all energies, as shown above.

That the globular cluster ULXs are different than most ULXs in star 
forming galaxies is not surprising. Globular cluster ULXs are essentially 
low-mass X-ray binaries (LMXBs) and nearly all field ULXs are high mass X-ray 
binaries (HMXBs). The binaries that make the globular cluster ULXs are also 
likely to be formed by dynamical interactions within the globular cluster  
\citep{2010ApJ...717..948I}, 
unlike the binary stellar evolution that makes field ULXs. As a result, the accretor in globular cluster ULXs is likely to be much older and thus not have the large magnetic fields often proposed for field ULXs. Differences in the donor stars will also be significant in globular cluster ULXs compared to field ULXs. Globular cluster sources will have old, low-mass donors while star forming ULXs typically have young, higher mass donors. This leads to differences in the orbital parameters and the composition of the accreted material, among others, which then may have implications for the resulting observables in the ULXs.

\begin{table*}
\centering
\caption{\label{table:all-lum} All luminosities in 0.2-10 keV. }
\begin{tabular}{lclclclclc}
\hline
\hline
Date     & Luminosity&  Lower Bound           & Upper  Bound & Obs \\ 
&(erg s$^{-1})$ &(erg s$^{-1})$ &(erg s$^{-1})$ \\
\hline
2000-03-19       &4.0$\times 10^{39}$   &  2.9$\times 10^{39}$  &5.6$\times 10^{39}$   & Chandra   \\ 
2000-06-12      &4.4$\times 10^{39}$ & 3.9$\times 10^{39}$   & 6.0$\times 10^{39}$  & Chandra    \\ 
2002-06-05  &4.9$\times 10^{39}$   &  4.4$\times 10^{39}$  &  5.5$\times 10^{39}$ & XMM \\ 
2004-01-01&1.0$\times 10^{39}$ &8.4$\times 10^{38}$&  1.3$\times 10^{39}$ & XMM \\ 
2008-02-23 & 4.7$\times 10^{39}$  & 3.3$\times 10^{39}$   & 7.0 $\times 10^{39}$  & Chandra \\
2010-02-27    &    7.4 $\times 10^{37}$   &        1.1$\times 10^{38}$  & 1.3 $\times 10^{38}$   & Chandra     \\ 
 2010-11-20    & 5.5$\times 10^{38}$ &   2.7$\times 10^{38}$  &1.2$\times 10^{39}$& Chandra \\ 
2011-02-14 & 2.1$\times 10^{39}$ &  1.8$\times 10^{39}$    &  2.5$\times 10^{39}$  & Chandra \\ 
 2011-02-21  & 1.1$\times 10^{39}$  &  8.7$\times 10^{38}$ &1.3$\times 10^{39}$   & Chandra  \\ 
 2014-08-04& 2.7$\times 10^{39}$  &2.0$\times 10^{39}$     &   3.3$\times 10^{39}$ & Chandra   \\ 
2015-02-24   &  6.3$\times 10^{37}$      &      9.8$\times 10^{36}$   &               2.74$\times 10^{38}$  & Chandra     \\ 
2016-01-05 & 3.1$\times 10^{39}$ &2.8$\times 10^{39}$&3.4$\times 10^{39}$ &XMM \\ 
2016-01-07 & 5.2$\times 10^{38}$&4.9$\times 10^{38}$&5.6$\times 10^{38}$& XMM \\ 
2016-01-09 &3.5$\times 10^{37}$ & 6.1$\times 10^{36}$ & 7.4$\times 10^{37}$ & XMM \\ 
2016-04-30  & 1.2$\times 10^{38}$  &2.0$\times 10^{37}$  &3.7$\times 10^{38}$ & Chandra \\ 

\hline
% SOURCE: /home/patricio/ast/esp01/analyses/ligtcurve/attempt_547/results.txt
\end{tabular}

\end{table*}

\section{CONCLUSIONS}
\label{sec:conclusions}

We confirm long-term variability in the X-ray emission
from the globular cluster black-hole candidate source XMMUJ122939.9+075333 in the extragalactic globular cluster
RZ2109. The system shows strong luminosity variability
over long and short time scales, dropping anywhere from as
low as $L_X \simeq 7 \times 10^{37}$ erg s$^{-1}$ to as high as $L_X \simeq 5 \times 10^{39}$ erg s$^{-1}$. The system also underwent significant changes in luminosity over very short time scales: observations four days
apart from each other showed a drop in luminosity by almost a factor of five.
Over 16 years of X-ray monitoring, the spectral shape remains extremely soft. The fitted \textit{Chandra} and \textit{XMM-Newton} spectra for the high count observations 
can be seen in the appendix. While the fit quality is too low
to make any statements in regards to spectral variability, it is
remarkable how consistently soft the spectra are.

\acknowledgments

KCD, SEZ, and MBP acknowledge support from Chandra grant GO4-15089A. 
SEZ and MBP also acknowledge support from the NASA ADAP grant NNX15AI71G.
This research has made use of data obtained from the Chandra Data Archive and the Chandra Source Catalog, and is based on observations obtained with XMM-Newton, an ESA science mission with instruments and contributions directly funded by ESA Member States and NASA.  We  also acknowledge use of NASA's Astrophysics Data System and Arxiv. 

\software
The following softwares and packages were used for analysis: \textsc{ciao}, software provided by the Chandra X-ray Center (CXC),   \textsc{heasoft} obtained from the High Energy Astrophysics Science Archive Research Center (HEASARC), a service of the Astrophysics Science Division at NASA/GSFC and of the Smithsonian Astrophysical Observatory's High Energy Astrophysics Division, \textsc{sas}, the XMM-Newton Science Analysis System, SAOImage DS9, developed by Smithsonian Astrophysical Observatory, \textsc{numpy} \citep{2011arXiv1102.1523V},  \textsc{matplotlib} \citep{2007CSE.....9...90H} and \textsc{astropy} \citep{2013A&A...558A..33A}.

\clearpage

\appendix

\begin{figure}[h]
\begin{tabular}{ll}
\includegraphics[scale=0.5]{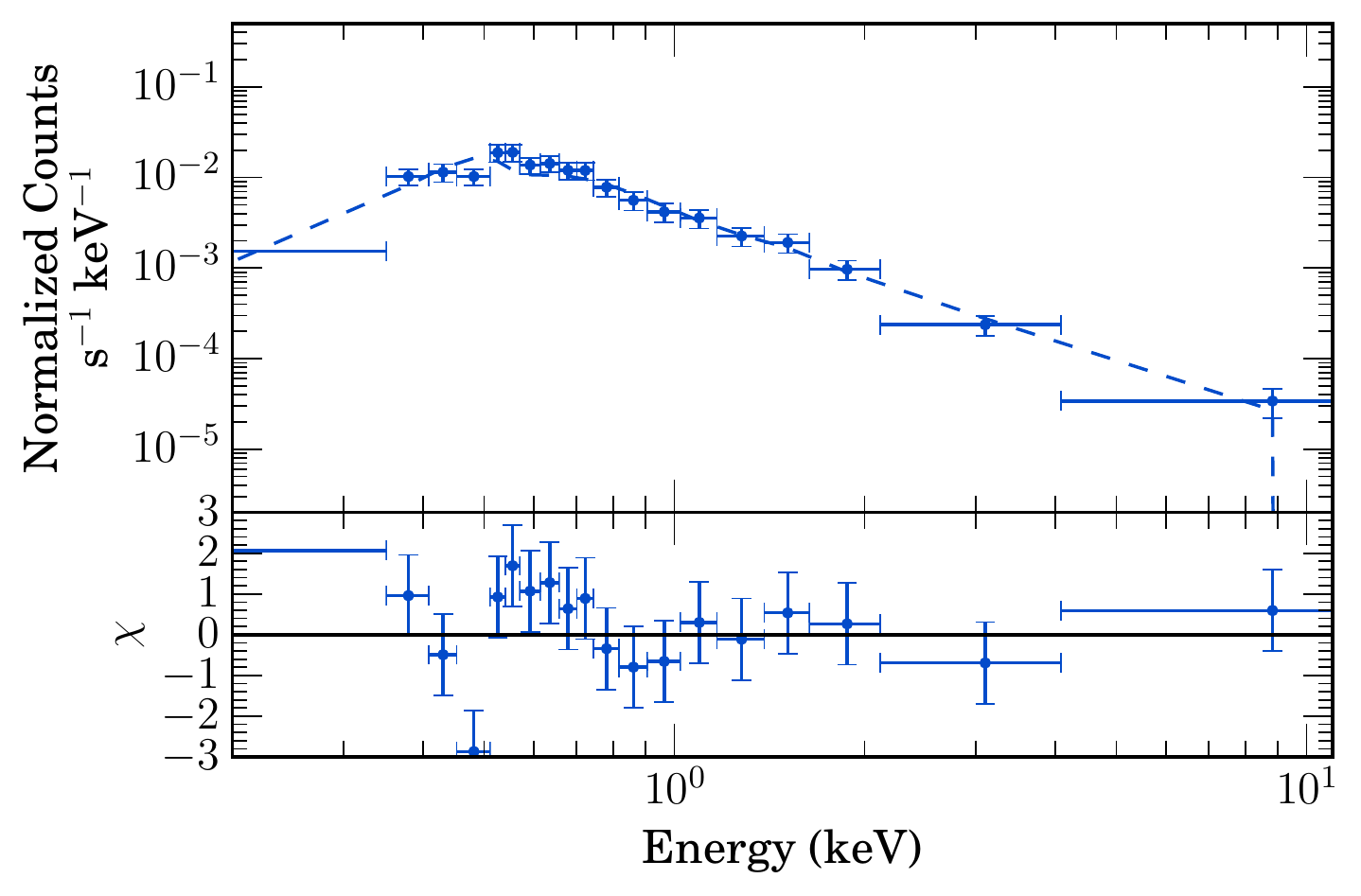}
&
\includegraphics[scale=0.5]{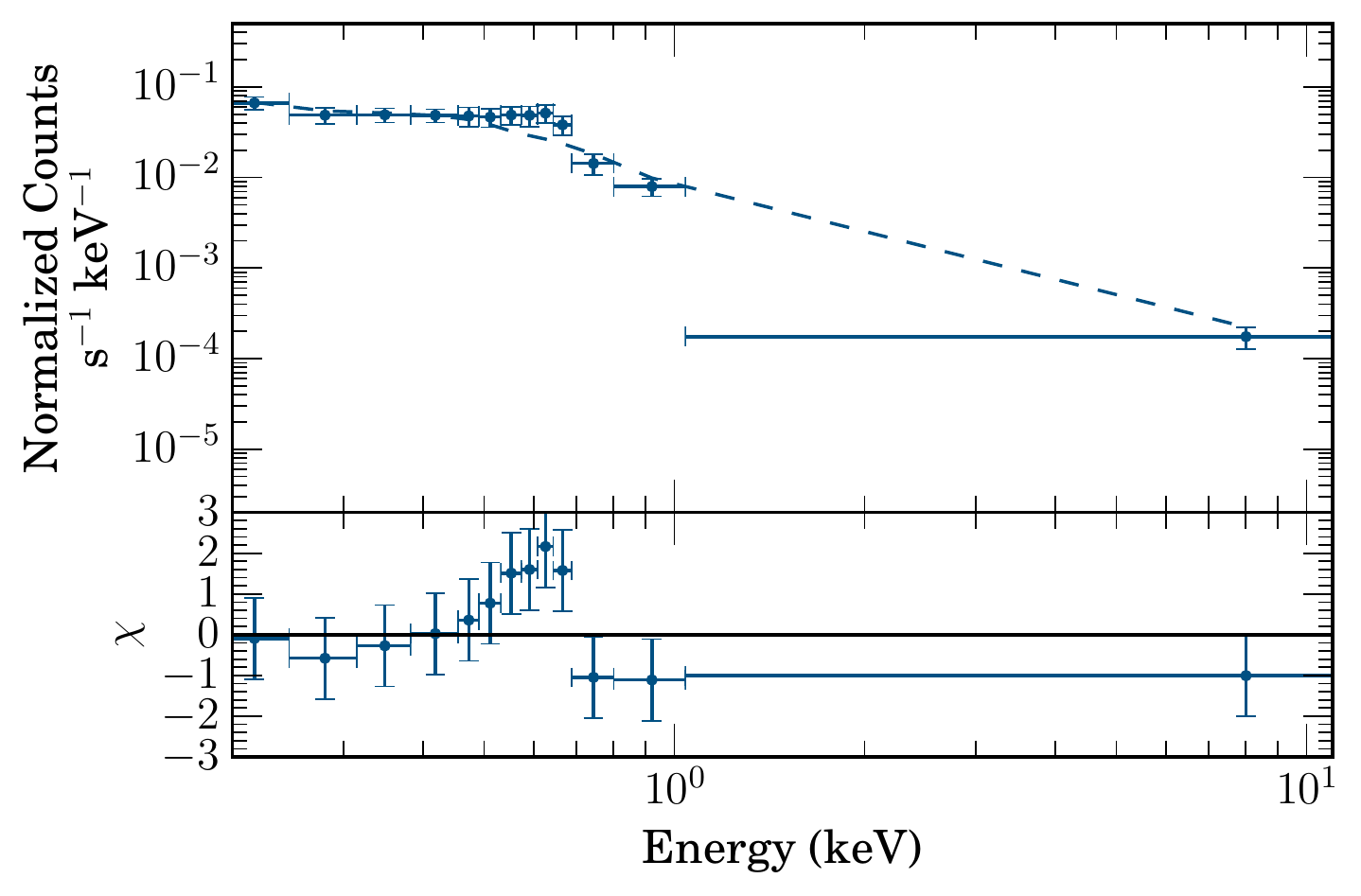}
\end{tabular}
\caption{Left:  Fitted spectrum with residuals of \textit{Chandra} ObsID 321 (2000-06-12). Right: Fitted spectrum with residuals of \textit{XMM-Newton} observation 0112550601 (2002-06-05, pn only).   }
\label{Fig:321_011}
\end{figure}

\begin{figure}[h]

\includegraphics[scale=0.5]{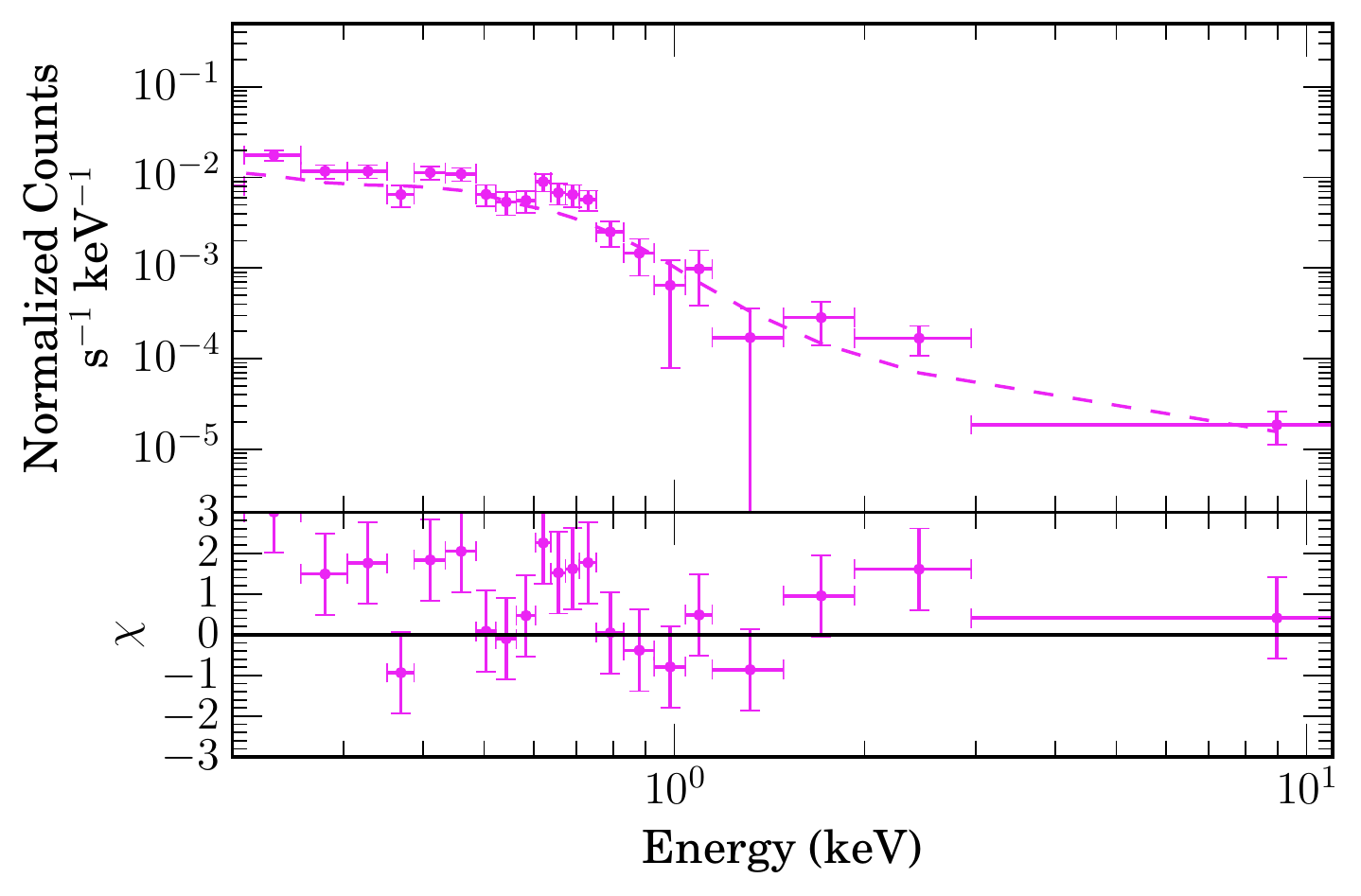}

\caption{Left:  Fitted spectrum with residuals of \textit{XMM-Newton} observation 0200130101 (2004-01-01, pn only).    }
\label{Fig:022}
\end{figure}

\begin{figure}[h]
\begin{tabular}{ll}
\includegraphics[scale=0.5]{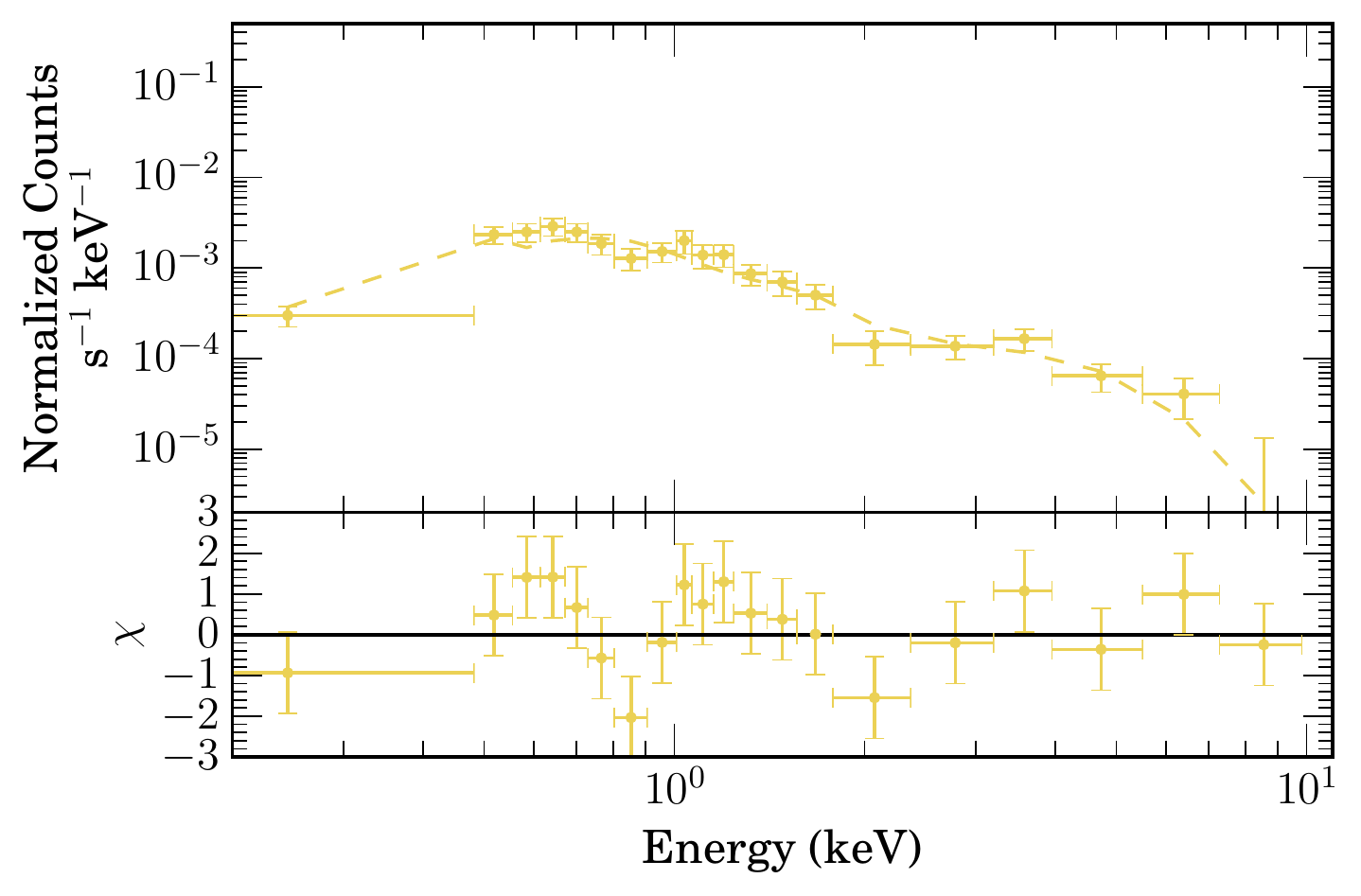}
&
\includegraphics[scale=0.5]{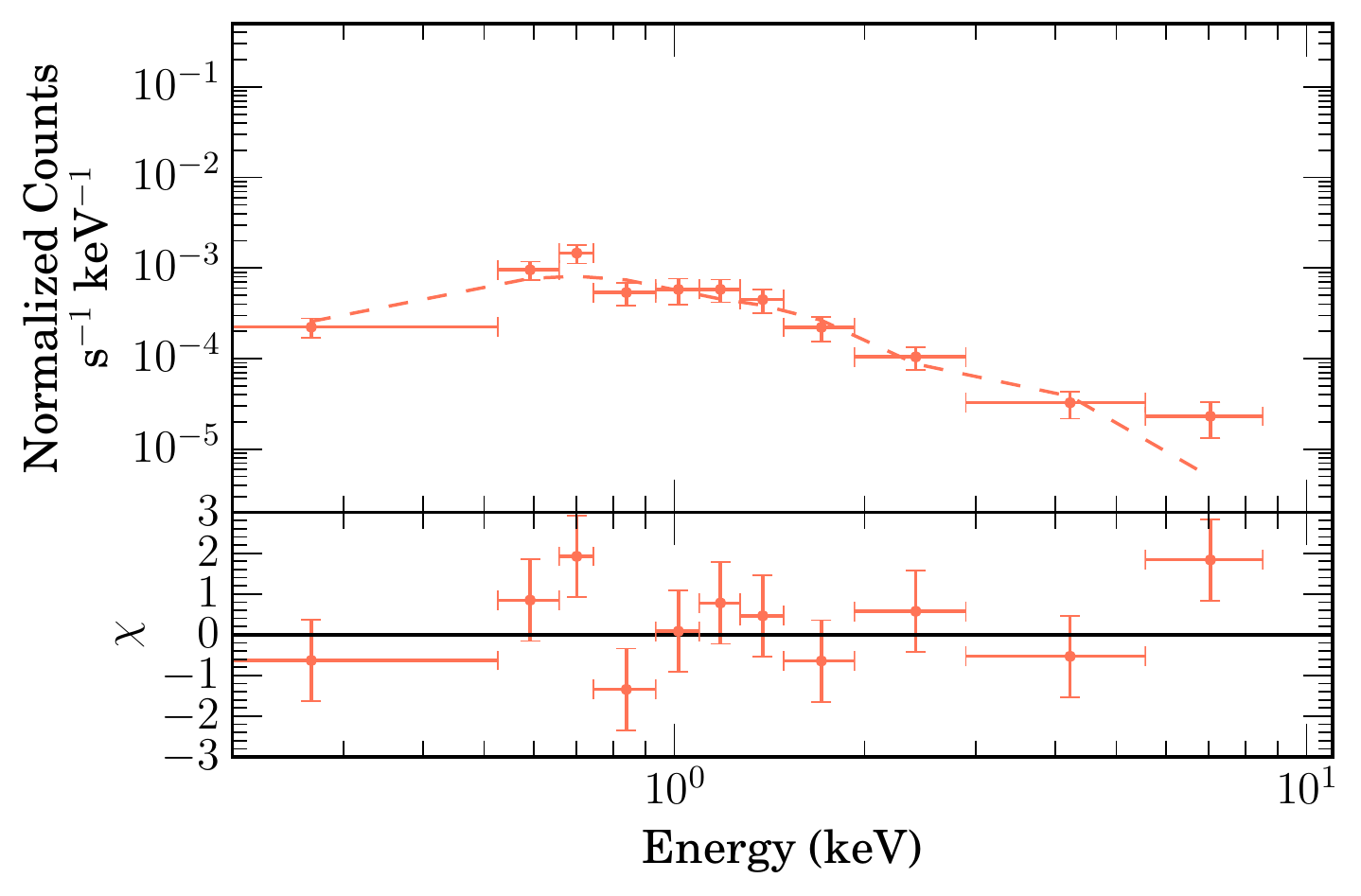}
\end{tabular}
\caption{Left: Fitted spectrum with residuals of \textit{Chandra} ObsID  12889 (2011-02-14). Right: Fitted spectrum with residuals of \textit{Chandra} ObsID 12888 (2011-02-21).  }
\label{Fig:322_321}
\end{figure}

\begin{figure}[h]
\begin{tabular}{ll}
\includegraphics[scale=0.5]{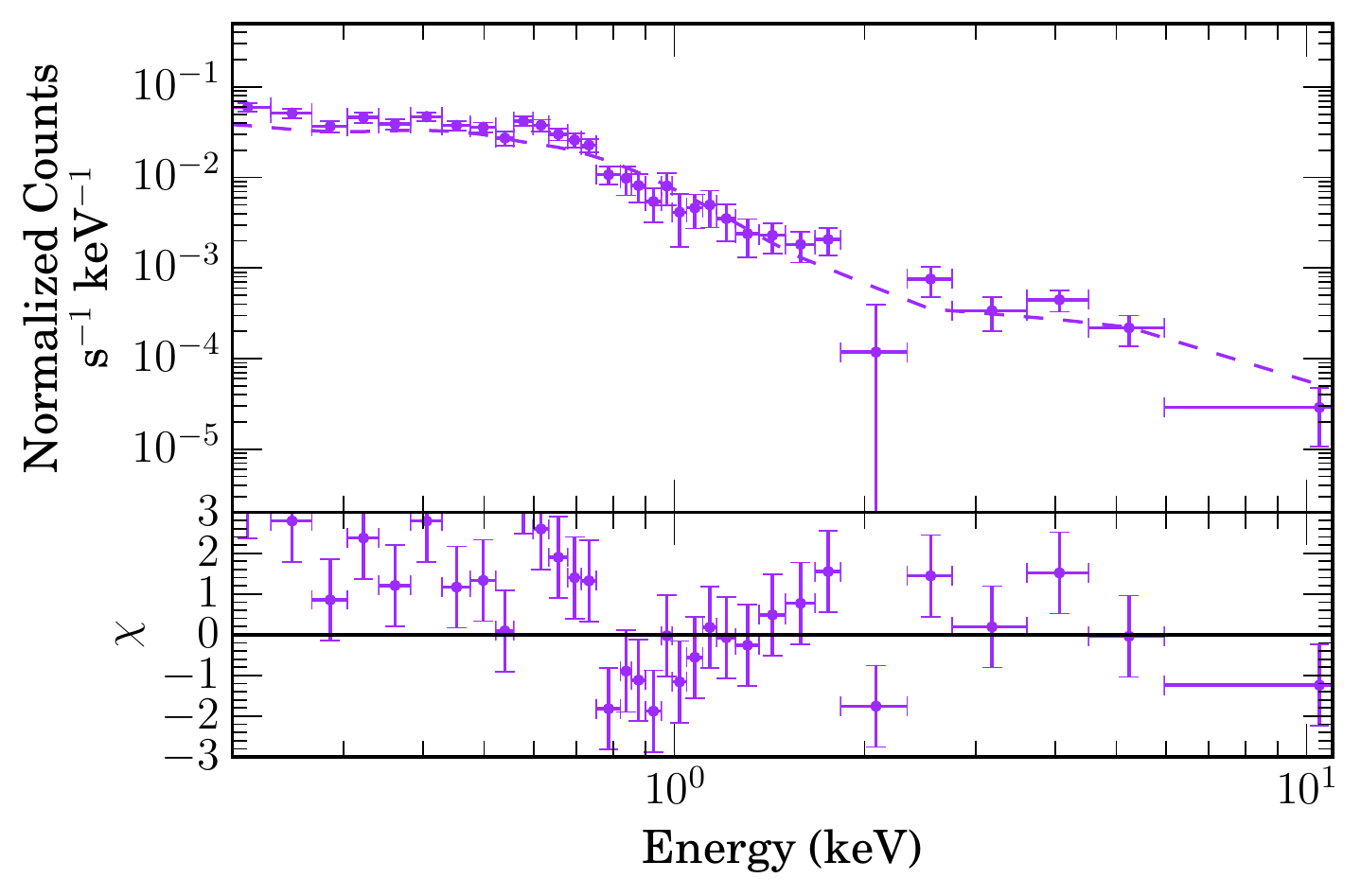}
&
\includegraphics[scale=0.5]{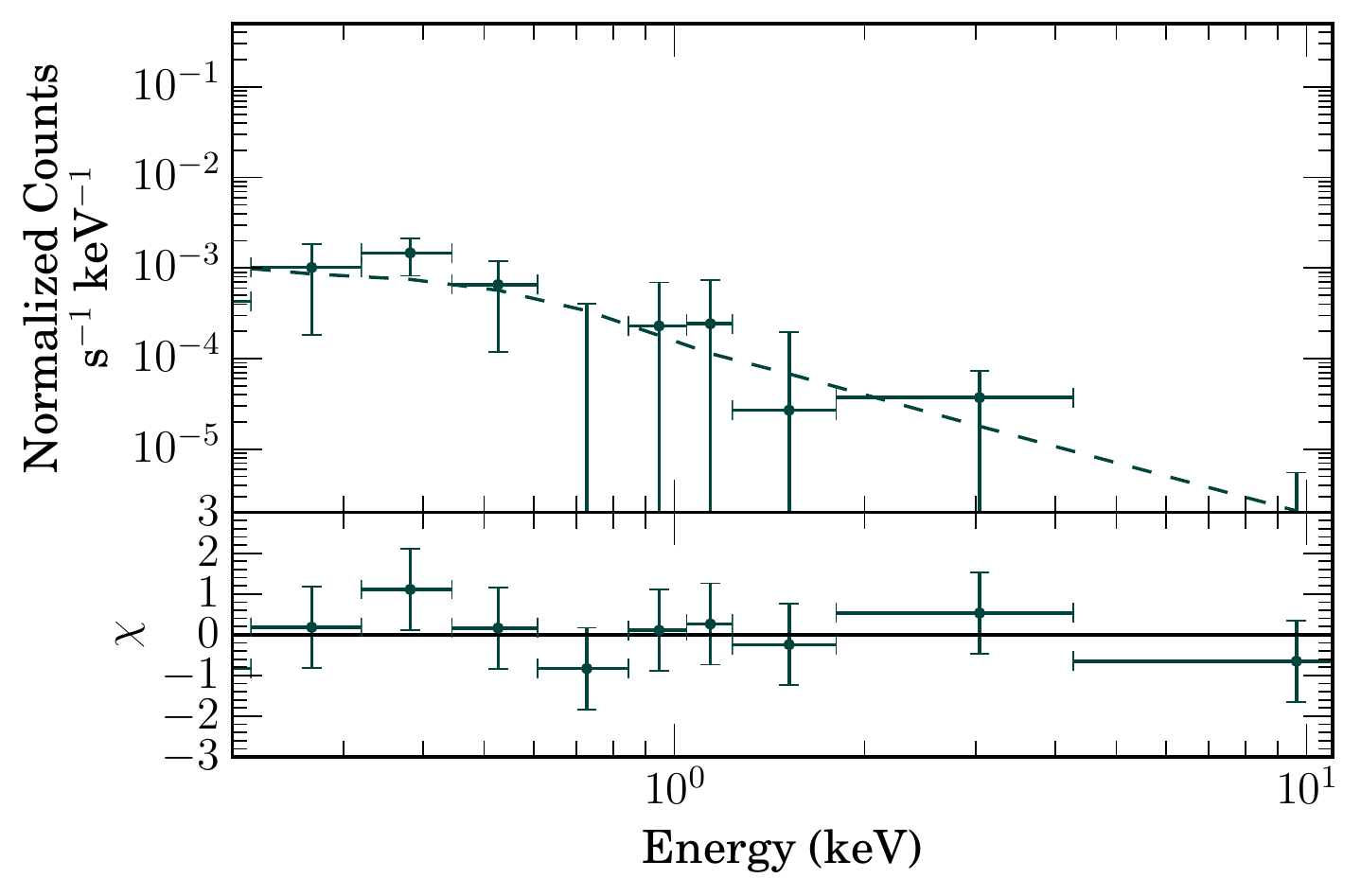}
\end{tabular}
\caption{Left: Fitted spectrum with residuals of \textit{XMM-Newton} observation 0761630101 (2016-01-05, pn only). Right: Fitted spectrum with residuals of \textit{XMM-Newton} observation 0761630301 (2016-01-09, pn only).   } 
\label{Fig:322_321}
\end{figure}

\clearpage
\bibliographystyle{apj}

\bibliography{rz2109}

\end{document}